\documentclass[12pt,a4paper]{article}
\usepackage[cp866]{inputenc}
\usepackage{amsmath}
\usepackage[dvips]{epsfig}
\usepackage{graphicx}
\usepackage{epsfig}
\newcommand{\be}{\begin{equation}}
\newcommand{\ee}{\end{equation}}
\newcommand{\ben}{\begin{equation*}}
\newcommand{\een}{\end{equation*}}
\newcommand{\bea}{\begin{eqnarray}}
\newcommand{\eea}{\end{eqnarray}}
\newcommand{\ar}{\begin{array}}
\newcommand{\arn}{\end{array}}

\def\pnot{\mbox{${\not{\hbox{\kern-3.0pt$p$}}}$}}
\def\qnot{\mbox{${\not{\hbox{\kern-2.0pt$q$}}}$}}
\def\enot{\mbox{${\not{\hbox{\kern-2.0pt$e$}}}$}}
\def\knot{\mbox{${\not{\hbox{\kern-2.0pt$k$}}}$}}

\def\fun#1#2{\lower3.6pt\vbox{\baselineskip0pt\lineskip.9pt\ialign
{$\mathsurround=0pt#1\hfil##\hfil$\crcr#2\crcr\sim\crcr}}}

\begin{document}
\numberwithin{equation}{section}     %\numerate of equation (first equation in each section)
\sloppy
\renewcommand{\baselinestretch}{1.0} %\constitute the line difference

\begin{titlepage}

\hskip 11cm \vbox{ \hbox{Budker INP 2020-10}  }
\vskip 3cm
\begin{center}
{\bf  Parton distributions in radiative corrections
	to the cross section of electron-proton scattering}
\end{center}

\centerline{V.S.~Fadin$^{a, b\,\dag}$, R.E.~Gerasimov$^{a, b\,\ddag}$}

\vskip .6cm

\centerline{\sl $^{a}$
  Budker Institute of Nuclear Physics of SB  RAS, 630090 Novosibirsk, Russia
}
\centerline{\sl $^{b}$
  Novosibirsk State University, 630090 Novosibirsk,
  Russia
}

\vskip 2cm

\begin{abstract}
The structure function approach and the parton picture,  developed for the theoretical description of the deep inelastic electron-proton scattering, also proved to be very effective for calculation of radiative corrections in Quantum Electrodynamics. We use  them  to calculate radiative corrections to the cross section of electron-proton scattering due to electron-photon interaction, in the experimental setup  with the recoil proton detection,  proposed  by A.A. Vorobev to  measure the  proton radius.   In the one-loop approximation,  explicit expressions for these corrections are obtained for arbitrary momentum transfers. It is shown that,  at  momentum transfers small compared with the proton mass,    various  contributions   to  the  corrections mutually cancel each other with power accuracy.  In two  loops,  the  corrections are obtained  in the leading logarithmic approximation.
\end{abstract}

%\vskip .5cm

\vfill \hrule \vskip.3cm \noindent $^{\ast}${\it Work supported
in part by the Ministry of  Science and Higher Education of Russian Federation and
in part by  RFBR,   grant 19-02-00690.} \vfill $
\begin{array}{ll} ^{\dag}\mbox{{\it e-mail address:}} &
\mbox{fadin@inp.nsk.su}\\
^{\ddag}\mbox{{\it e-mail address:}} &
\mbox{r.e.gerasimov@inp.nsk.su}\\
\end{array}
$
\end{titlepage}

\vfill \eject

\section{Introduction}
Although after  appearance of the paper \cite{Beyer:2017gug} the  "proton radius puzzle" \cite{Pohl:2013yb, Carlson:2015jba} --
the striking  difference in the proton radius values extracted  from  the 2S-2P transition in muonic hydrogen \cite{Pohl:2010zza, Antognini:1900ns} and obtained from electron-proton  scattering and hydrogen spectroscopy \cite{Mohr:2008fa}  (for a review, see Ref.~\cite{Bernauer:2014nqa}) -- seems partially (regarding the contradiction between the results of experiments with muonic  and usual  hydrogen) resolved,   the contradiction  between  muonic  hydrogen and electron-proton  scattering results remains.   Moreover,  latest electron scattering  experiments at Jlab~\cite{Zhan:2011ji} and MAMI~\cite{Bernauer:2010wm} and  hydrogen spectroscopy experiments \cite{Beyer:2017, Fleurbaey:2018fih} not only did not resolve the puzzle, but made it even more confusing.

Currently new scattering experiments are being prepared. A distinctive feature of one of them \cite{Vorobev}, which was suggested by A.A. Vorobev and  has  to  be performed with a low-intensity electron beam at MAMI,  is that instead of detecting a scattered electron, as in previous experiments, it is supposed to detect with a high precision a recoil proton in the region of low momentum transfers ($0.04\;\mathrm{ GeV}^2> Q^2> 0.001 \;\mathrm{ GeV}^2$). The aim is to extract the proton radius with 0.6 percent precision, which could be decisive in solving the proton radius puzzle. To this end, it is planned to achieve 0.2 percent accuracy of the cross section $d\sigma/(d Q^2)$ measurement.

Such accuracy requires precise  account of radiative corrections. Although calculation of the radiative corrections to the electron-proton scattering cross section has a long history (see, for example, Refs.~\cite{Tsai:1961zz}-\cite{Maximon:2000hm}\footnote{higher
	order corrections to the lepton line was considered for the standard
	experimental set-up with scattered electron measurement in
	\cite{Arbuzov:2015vba}.}, and recent
reviews \cite{Carlson:2007sp}-\cite{Afanasev:2017gsk})  the results obtained before cannot be completely applied to the experiment discussed above. The reason it that they were obtained for experiments in which scattered electrons were detected (honestly speaking, there was the experiment \cite{Qattan:2004ht} where the recoil proton was detected; but calculation  of the radiative corrections to this experiment was not explained). Since the radiative corrections include contributions of inelastic processes with photon emission, they depend  strongly  on experimental conditions, so that the corrections calculated  for experiments with detection of scattered electrons are not suitable for experiments with detection  of recoil proton. It occurs \cite{Fadin:2018dwp} that the radiative corrections for experiments with detection  of recoil proton have a new unexpected and pleasant property  -- cancellation of the most important corrections, which are  due to electron-photon interaction\footnote{Cancellation of leptonic radiative corrections to deep
	inelastic scattering was discussed in \cite{Akhundov:1994my} and
	\cite{Arbuzov:1995id}.},  in the region of low  momentum transfers. In \cite{Fadin:2018dwp},  the cancellation  of not only infrared, but also collinear singularities  was shown and  a simple physical explanation of this phenomenon was given.
It was also argued that in the one-loop approximation the accuracy of the cancellation   is higher than the logarithmic, and  the terms not having the collinear singularities  (constant terms) are cancelled as well.

Here we refine  the results of \cite{Fadin:2018dwp} and  get new ones, with a wider scope of applicability, using the structure function approach  and the parton picture,  developed for the theoretical description of the deep inelastic electron-proton scattering \cite{Gribov:1972ri}-\cite{Dokshitzer:1977sg}  and adopted  in \cite{Kuraev:1985wb} for calculation of radiative corrections in QED.

\section{Statement of the  approach}
Following \cite{Fadin:2018dwp}, we  denote four-momenta of initial and final electron (proton)  as $l\;$($p$) and  $l'\;$($ p'$); $\;\;l^2=l'^2=m^2, \;\; p^2=p'^2=M^2$, and  use the designations $Q^2 = - q^2, \;\; q=p-p'$  both for elastic the and inelastic processes.

The cross section of electron-proton scattering with radiative corrections  due only to electron interaction  can be considered as inclusive proton-electron scattering cross section.
It means that it  can be  written as
\be
E'_p\frac{d^3\sigma}{d^3p'}
=\frac{(\alpha(Q^2))^2}{Q^4}\frac{1}{\sqrt{(pl)^2-m^2M^2}}J^{\mu\nu}(p, p')
W_{\mu\nu}(l, q)~, \label{cross section = L munu W munu}
\ee
where
\be
\alpha(Q^2) = \frac{\alpha}{1-{\cal P}(q^2)}~,
\ee
${\cal P}(q^2)$ is the vacuum polarisation, which is real at $q^2=-Q^2 <0$;
$J^{\mu\nu}(p, p')$ is the proton current tensor
\be
J^{\mu\nu}(p, p') = \overline{\sum}_{pol}J^\mu J^{*\nu}~,
\ee
 $\overline{\sum}_{pol}$  means summation over final polarisations and averaging over initial ones,
 \be
 J^\mu = \bar{u}(p')\,\left(f_1(Q^2)\,\gamma^\mu + f_2(Q^2)\,\frac{[\gamma^\mu, \gamma^\nu]q_\nu}{4M}\right)\,u(p)~,
 \ee
$f_1(Q^2)$ and  $f_2(Q^2)$ are the   the Dirac and Pauli form factors of the proton,  and $W_{\mu\nu}(l, q)$ is the deep inelastic scattering tensor,
 \be
 {W}_{\mu\nu}(l, q)=\frac{1}{4\pi} \overline{\sum}_X\langle
 l|j_\nu^{(e)}(0)|X\rangle~\langle
 X|j_\mu^{(e)}(0)|l\rangle~(2\pi)^4\delta(q+l-p_X)~. \label{W mu nu}
\ee
Here $|l\rangle$ is the initial electron state, $|X\rangle$ is any state which can be produced in photon-electron collisions, $\overline{\sum}_X$ means averaging over initial electron polarisations and   summation over discrete and integration over continuous variables of $|X\rangle$,
$j_\mu^{(e)}$ is the electron electromagnetic current   operator.

Taking into account  conservation of the current, one can represent $W_{\mu\nu}$ in the form
\be
{W}^{\mu\nu}(l, q)=-\left(g^{\mu\nu}-\frac{q^\mu q^\nu}{q^2}\right)F_1(x, Q^2)+\frac{1}{(l q)}\left(l^{\mu}-\frac{(l q)}{q^2}q^\mu\right)\left(l^{\nu}-\frac{(lq)}{q^2 }q^\nu\right)F_2(x, Q^2)~, \label{Wmunu through Fi}
\ee
where
\be
x=Q^2/(2(l q))
\ee
is the Bjorken variable  and $F_i(x, Q^2)$ are   the electron structure functions. They are expressed in terms of the convolutions $W_i$ of  the tensor ${W}^{\mu\nu}(l, q)$
\be
W_g=W_{\mu\nu}(l, q) g^{\mu\nu}~, \;\; W_{l}=W_{\mu\nu}(l, q) l^\mu l^\nu~  \label{Wi}
\ee
using the relations
\be
F_1(x, Q^2) = \frac12\left(\frac{Q^2\;W_l}{Q^2 m^2+(ql)^2}-W_g\right)~, \label{F1}
\ee
\be
F_2(x, Q^2) = \frac12\frac{Q^2 (ql)}{(Q^2 m^2+(ql)^2)}\left(\frac{3Q^2\;W_l}{Q^2 m^2+(ql)^2}-W_g\right)~. \label{F2}
\ee
Calculating the tensor $J^{\mu\nu}$,
\be
J^{\mu\nu} =G_M^2(Q^2)\left(g_{\mu\nu}q^2-{q_\mu q_\nu}\right)+\frac{Q^2G_M^2(Q^2) +4M^2 G_E^2(Q^2)}{4M^2+Q^2}P^\mu P^\nu~,
\ee
where $P =p+p'$, $G_E(Q^2)$ and $G_M(Q^2)$ are the  proton electric and magnetic form factors,
\be
G_M(Q^2)=f_1(Q^2)+f_2(Q^2),\;\; G_E(Q^2)=f_1(Q^2)-\frac{Q^2}{4M^2}f_2(Q^2)~,
\ee
performing tensor convolution and using
\be
\frac{d^3p'}{2E'_p} =\frac{\pi}{4}\frac{Q^2 dQ^2 dx}{x^2\sqrt{(pl)^2-m^2M^2}}~,
\ee
we obtain
\[
\frac{d\sigma}{dQ^2 dx} = \frac{\pi(\alpha(Q^2))^2}{2x^2Q^2((pl)^2-m^2M^2)}
\left[(2Q^2G_M^2 -4M^2G_E^2)F_1(x, Q^2) +\right.
\]
\be\left.
 \left(-G_M^2(m^2 Q^2 +(ql)^2)+\frac{Q^2G_M^2 +4M^2G_E^2}{4M^2+Q^2}(Pl)^2  \right)\frac{F_2(x, Q^2)}{(ql)} \right]~.  \label{d sigma d q2 dx}
\ee
Here
\be
(ql) =\frac{Q^2}{2x}~,\;\;   (Pl) = 2(pl) - \frac{Q^2}{2x}~, \;\; (pl)= M E_l~,
\ee
where  $E_l$ is  the energy of the incident electron in the rest frame of the initial proton.

In the Born approximation, the cross section $\frac{d\sigma_{B}}{dQ^2}$ is determined    by (\ref{d sigma d q2 dx}) with $\alpha(Q^2) =\alpha, \; F_2(x, Q^2)=2F_1(x, Q^2) =\delta(1-x)$:
\be
\frac{d\sigma_{B}}{dQ^2} = \frac{\pi\alpha^2 M^2}{Q^4}
\frac{(4(pl) -Q^2)^2 + (Q^2+4M^2)(Q^2-4m^2)}{(Q^2+4M^2)((pl)^2-m^2M^2)}(\epsilon G_E^2 + \tau G_M^2)
~, \label{d sigma B}
\ee
\be
\epsilon = \frac{(4(pl) -Q^2)^2 - Q^2(Q^2+4M^2)}{(4(pl) -Q^2)^2 + (Q^2+4M^2)(Q^2-4m^2)}~, \;\; \tau =\frac{Q^2}{4M^2}~.
\ee
Formula (\ref{d sigma d q2 dx}) gives  the  exact expression for the cross section of  electron-proton  scattering  taking into account all  processes   of electron-photon interaction. The radiation correction due to this interaction is determined by the equation
\be
\delta_{e\gamma} =\frac{\int_0^1 dx \; \frac{d\sigma}{dQ^2 dx} } {\frac{d\sigma_B}{dQ^2}} -1  ~ \label{delta el}
\ee
and can be written as
\be
\delta_{e\gamma} =\frac{1+\delta^e} {(1-{\cal P}(q^2))^2} -1  ~,  \label{delta e}
\ee
where $\delta^e$ is the correction  associated with the electron structure, that is, with the difference  $F_2(x, Q^2) $ and $2 F_1(x, Q^2) $ from  $\delta(1-x)$.

Our main goal here is to calculate just this  correction. As for he vacuum polarisation ${\cal P}(q^2)$,  it is  well known and  we have nothing new to say about it. For completeness, we provide the necessary information in Appendix A.

In the proposed experiments to measure the proton radius, the momentum transfers are large compared to the electron mass,  $Q^2 \gg m^2$. Below, we will be mainly interested in this particular area.
Here,  it is  convenient to use the following representation of the  cross sections (\ref{d sigma d q2 dx})
and (\ref{d sigma B})
\be
\frac{d\sigma}{dQ^2 dx} = \frac{4\pi(\alpha(Q^2))^2}{Q^4}\bigg[\frac{F_2(x, Q^2)}{x}R(x, Q^2) +\frac{Q^2(2Q^2G_M^2 -4M^2G_E^2)}{8x^2(pl)^2}\big(F_1(x, Q^2)-\frac{F_2(x, Q^2)}{2x} \big)\bigg]~, \label{d sigma dQ^2 dx}
\ee
where
\be
R(x, Q^2) = \bigg(1-\frac{Q^2}{2x(lp)}\bigg)\frac{{Q^2}G_M^2 +4M^2G_E^2}{4M^2+Q^2} + \frac{Q^2}{8x^2(pl)^2} \frac{Q^2(2M^2 + Q^2)G_M^2 -8M^4G_E^2}{4M^2+Q^2}~, \label{Rx}
\ee
and
\be
\frac{d\sigma_B}{dQ^2} = \frac{4\pi \alpha^2}{Q^4}R(1, Q^2)~. \label{d sigma B d q 2}
\ee

\section{Elastic scattering}

For the elastic scattering, when  $|X\rangle$ in (\ref{W mu nu}) are  the one-electron states with the momentum $l'$, we have
\be
\langle X|j_\mu^{(e)}(0)|l\rangle =\langle l'|j_\mu^{(e)}(0)|l\rangle  =
\bar{u}(l')\,\left(f^e_1(Q^2)\,\gamma_\mu - f^e_2(Q^2)\,\frac{[\gamma_\mu, \gamma^\nu]q_\nu}{4m}\right)\,u(l)~,\label{vertex}
\ee
where $f^e_i(Q^2)$  are the electron form factors.  Using  (\ref{F1}), (\ref{F2}) and (\ref{Wi}),
one obtains for the elastic contributions $F^{el}_i$  to the electron structure functions $F_i$
\[
F^{el}_1(x, Q^2)= \frac{1}{2}\delta(1-x)\big(f^e_1(Q^2)+f^e_2(Q^2)\big)^2,
\;\;
\]
\be
F^{el}_2(x, Q^2)=\delta(1-x)\Big[\big(f^e_1(Q^2)\big)^2+\frac{Q^2}{4m^2} \big(f^e_2(Q^2)\big)^2\Big]~.
\label{Fi in terms of fi}
\ee
Eq. (\ref{d sigma d q2 dx}) then gives
\[
\frac{d\sigma^{el}}{dQ^2} = \frac{\pi(\alpha(Q^2))^2}{4Q^2((pl)^2-m^2M^2)}
\Big[(4(pl) -Q^2)^2 \frac{(Q^2G_M^2 +4M^2G_E^2)(Q^2g_M^2 +4m^2g_E^2)}{Q^2(Q^2+4M^2)(Q^2+4m^2)}
\]
\be
+ \big(Q^2G_M^2 -4M^2G_E^2\big)g_M^2-4m^2G_M^2g_E^2]\Big]~, \label{d sigma el dQ2}
\ee
where
\be
g_M=f^e_1(Q^2)+f^e_2(Q^2),\;\; g_E=f^e_1(Q^2)-\frac{Q^2}{4m^2}f^e_2(Q^2)~. \label{d sigma el}
\ee
Formally, Eq.  (\ref{d sigma el dQ2}) gives  the exact  expression for the  cross section of elastic  electron-proton  scattering with one-photon exchange.
But essentially it has no physical meaning due to the infrared singularity. Taking into account all terms of the expansion in terms of the coupling constant $\alpha$ makes it zero, and each term of the expansion requires the regularization of this singularity.  If the infrared divergency is regularised by the photon mass $\lambda$,  in the one-loop approximation one has \cite{BLP}
\be
f^e_1(Q^2) = 1 -\frac{\alpha}{\pi \beta}\bigg[
\big(\ln\xi -\beta\big)\big(\ln\frac{m}{\lambda}-1\big) -\frac14\ln^2 \xi +\ln \xi \ln(1+\xi)+\frac{\pi^2}{12}
+ \text{Li}_2(-\xi)
+ \frac{\beta(\xi-1)}{\xi+1} \ln\xi\bigg] ~, \label{one-loop f1e}
\ee
\be
f^e_2(Q^2) = \frac{\alpha}{2\pi}
\frac{\sqrt{1-\beta^2}}{\beta}
\ln\xi
~,\label{one-loop f2e}
\ee
where  $\beta$  is the  velocity  of one of the electrons in the rest frame of the other,
\be
\beta = \frac{\sqrt{Q^2(Q^2+4m^2)}}{Q^2+2m^2}~,\;\; \xi=\sqrt{\frac{1+\beta}{1-\beta}}~,\;\; \text{Li}_2(x) =-\int_0^x\frac{dt}{t}\ln(1-t)~. \label{beta}
\ee

The part $\delta^e_{vertex}$ of the  associated with the electron structure correction $\delta^e$ introduced by elastic scattering is determined by the difference between the vertex (\ref{vertex}) and the Born one, in which $f_1^e(Q^2) =1$ and   $f_2^e(Q^2)=0$. At  \( Q^2\gg m^2 \)  the   Pauli form factor $f_2^e(Q^2) $ is suppressed by a power-law, $f_2^e(Q^2) \sim m^2/Q^2$, so that in the one-loop approximation we have
\be
\delta_{vertex}^{e} = 2 \big(f^{e}_1(Q^2) -1\big) = \frac{\alpha}{\pi}\left[
-\left(\ln\frac{Q^2}{m^2} - 1\right) \ln \frac{m^2}{\lambda^2}
-\frac{1}{2}\ln^2\frac{Q^2}{m^2}
+\frac{3}{2}\ln\frac{Q^2}{m^2}
+\frac{\pi^2}{6}
-2
\right]
~. \label{delta vertex}
\ee

\section{One photon emission}
For one photon emission, when   the states  $|X\rangle$ in (\ref{W mu nu}) are states of an electron with momentum  $l'$ and photon with momentum $k$, we have
\be
W_{\mu\nu}(l, q) = -\frac{e^2}{8\pi}\int  K_{\mu\nu}(2\pi)^4\delta^{(4)}(q+l-l'-k)\frac{d^3l'}{(2\pi)^32E'_l}\frac{d^3k}{(2\pi)^3 2\omega}~, \label{W mu nu  1}
\ee
where
\[
K_{\mu\nu} = g^{\rho\sigma}tr[(\hat{l'}+m)L_{\mu\rho}(\hat{l}+m)\gamma^0L^\dag_{\nu\sigma}\gamma^0]~,
\]
\be
L_{\mu\rho} =
\gamma_\mu\frac{\hat{l}-\hat{k}+m}{-2\kappa}\gamma_\rho+\gamma_\rho\frac{\hat{l'}+\hat{k}+m}{2\kappa'}\gamma_\mu~, \;
\kappa=(kl)~,\;\;\kappa'=(kl')= Q^2\frac{1-x}{2x}~. \label{K mu nu}
\ee
Moving on to integration over $\kappa$, we obtain from (\ref{W mu nu  1}) for the convolutions (\ref{Wi}) of $W_{\mu\nu}(l, q)$
\be
W_i(lq, q^2)
=-\frac{\alpha}{8\pi}\int_{\kappa_-}^{\kappa_+}\frac{d\kappa}{\sqrt{I_C}}A_i~, \label{Wi 1}
\ee
where
\be
I_C = (Q^2+2\kappa')^2+4m^2Q^2~, \;\; A_g = K_{\mu\nu} g^{\mu\nu}~, \;\; A_l=K_{\mu\nu} l^\mu l^\nu~, \label{Ai}
\ee
and
\be
{\kappa_{\pm}}= \frac{\kappa'}{2(m^2+2\kappa')}\left( 2m^2+2\kappa'+Q^2 \pm \sqrt{I_C} \right)~.
\ee
The integration limit $\kappa_-$ ($\kappa_+$) corresponds to forward (backward) virtual  Compton scattering.

Direct calculation of the convolutions  $A_i$ (\ref{Ai}) with use of (\ref{K mu nu}) gives
\be
A_g = 4\left(\frac{m^2}{\kappa^{\prime \; 2}} +\frac{m^2}{\kappa^{ 2}} -\frac{2m^2+Q^2}{\kappa \kappa^{\prime} } +\frac{2}{\kappa'}-\frac{2}{\kappa}\right)(2m^2-Q^2) +8\left(\frac{\kappa}{\kappa'}+\frac{\kappa'}{\kappa}\right)~,
\ee
\[
A_l = 2m^2\Bigg[\left(\frac{m^2}{\kappa^{\prime \; 2}} +\frac{m^2}{\kappa^{ 2}} -\frac{2m^2+Q^2}{\kappa \kappa^{\prime} } +\frac{2}{\kappa'}-\frac{6}{\kappa}\right)(4m^2+Q^2)
\]
\be
 +\frac{4m^2 \kappa'}{\kappa^2} +\frac{2(2m^2 +\kappa)}{\kappa'}+\frac{2(4m^2 -3\kappa')}{\kappa} +4\Bigg] -4(Q^2+2\kappa'-2\kappa)~.
\ee
Note that $A_i$ should be obtained from  Eqs.(7.39)  of \cite{Baier:1973} with the substitutions
\be
w^2\rightarrow m^2+2\kappa'~, \;\; q^2\rightarrow m^2-2\kappa~, \;\; \Delta^2\rightarrow -Q^2~,\;\;\Delta_1^2\rightarrow 0~. \;\;
\ee
Unfortunately, in the expression for $A^{(\frac12)}_2$ in Eqs.(7.39)   there is a  misprint;
it contains the extra term ${8(p_1p_2)m^4}/{(q^2-m^2)^2}$.

Calculation of $W_i(lq, q^2)$  (\ref{Wi 1})  is performed using the integrals
\[
\int_{\kappa_-}^{\kappa_+}\frac{d\kappa}{\sqrt{I_C}} =\frac{\kappa'}{m^2+2\kappa'}~, \;\;
\int_{\kappa_-}^{\kappa_+}\frac{\kappa d\kappa}{\sqrt{I_C}} =\frac{\kappa'^2(2m^2+2\kappa'+Q^2)}{2(m^2+2\kappa')^2}~,
\]
\be
\int_{\kappa_-}^{\kappa_+}\frac{d\kappa}{\kappa^2\sqrt{I_C}} =\frac{1}{m^2\kappa'}~, \;\; \int_{\kappa_-}^{\kappa_+}\frac{d\kappa}{\kappa  \sqrt{I_C}} ={\cal L}~,  \label{int dkappa}
\ee
where
\be
{\cal L} =\frac{1}{\sqrt{(Q^2+2\kappa')^2+4m^2Q^2}}  \ln\left(\frac{2m^2+2\kappa'+Q^2+\sqrt{I_C}}{2m^2+2\kappa'+Q^2-\sqrt{I_C}}\right)~.
\ee
It gives
\[
W_g= -\frac{\alpha}{2\pi}\left[(Q^2-2m^2)\Big[\Big(\frac{Q^2+2m^2}{\kappa'}+2\Big){\cal L} -\Big(\frac{m^2}{\kappa'^2} +\frac{2}{\kappa'}\Big)\frac{\kappa'}{m^2+2\kappa'} -\frac{1}{\kappa'}\Big]
\right.
\]
\be
\left. +\frac{\kappa'(2m^2+2\kappa'+Q^2)}{(m^2+2\kappa')^2} +2\kappa' {\cal L}\right]~, \label{W1}
\ee
\[
W_l= -\frac{\alpha}{2\pi}\Bigg[\frac{m^2}{2}\Bigg(8m^2-6\kappa'-(Q^2+4m^2)\Big(\frac{Q^2+2m^2}{\kappa'}+6\Big)\Bigg){\cal L} +m^2\Big(2+\frac{Q^2+4m^2}{2\kappa'}\Big)
\]
\be
 +\Bigg(m^2\Big((Q^2+4m^2)(\frac{m^2}{\kappa'^2} +\frac{2}{\kappa'})+\frac{4m^2}{\kappa'} +6\Big) -Q^2-2\kappa'\Bigg)\frac{\kappa'}{2(m^2+2\kappa')} \Bigg]~.\label{W2}
\ee

The region of variation of $x$ at fixed $Q^2$ is determined by the conditions $M_X^2 =(m^2+2\kappa') \ge m^2$ and $(lq) \le E_lq_0 +\sqrt{E_l^2-m^2}\sqrt{q_0^2+Q^2}$   with $q_0=M-E'_p = -Q^2/(2M)$, i.e
\be
1\ge x\ge x_-~,\;\;   x_-=\frac{MQ^2}{\sqrt{E_l^2-m^2} \sqrt{Q^2(4M^2 +Q^2)} -E_l Q^2}~.
\ee
But  expressions (\ref{W1}) and (\ref{W2}) can not be used arbitrarily close to $x=1$ (i.e. for sufficiently small $\kappa'$)  because of the infrared divergency. The divergency must  be regularised in the same way as in the vertex correction (\ref{delta vertex}), i.e. by the photon mass $\lambda$. Taking into account the photon mass changes both the measure and the limits of integration in (\ref{Wi 1}):
\be
I_C \rightarrow I_C(\lambda)  = (Q^2+2\kappa' +\lambda^2)^2+4m^2Q^2~, \label{IClambda}
\ee
\be
{\kappa_{\pm}}\rightarrow ={\kappa_{\pm}}(\lambda) = \frac{ (\kappa'+\lambda^2)(2m^2+2\kappa'+Q^2+\lambda^2) \pm \sqrt{(\kappa'^{\;2} -m^2\lambda^2)I_C(\lambda)} }{2(m^2+2\kappa'+\lambda^2)}~.  \label{kappapmlambda}
\ee
In the region $m^2 \gg \kappa'> m\lambda$ at $\lambda \rightarrow 0$ they can be taken as
\be
I_0   = Q^2(Q^2+4m^2)~, \label{I0}
\ee
\be
{\kappa^0_{\pm}} = \frac{\left( \kappa'(2m^2+Q^2) \pm \sqrt{(\kappa'^{\;2} -m^2\lambda^2)I_0} \right)}{2m^2}~.  \label{kappapm0}
\ee
The  singular terms in $A_i$ are
\be
A_g = 4(2m^2-Q^2)\left(\frac{m^2}{\kappa^{\prime \; 2}} +\frac{m^2}{\kappa^{ 2}} -\frac{2m^2+Q^2}{\kappa \kappa^{\prime} }\right)~,
\ee
\be
A_l = 2m^2(4m^2+Q^2)\left(\frac{m^2}{\kappa^{\prime \; 2}} +\frac{m^2}{\kappa^{ 2}} -\frac{2m^2+Q^2}{\kappa \kappa^{\prime} }\right) ~.
\ee
Corresponding   integrals become
\[
\int_{\kappa^0_-}^{\kappa^0_+}\frac{d\kappa}{\sqrt{I_0}} =\frac{\sqrt{(\kappa'^{\;2} -m^2\lambda^2)}}{m^2}~, \;\;
\int_{\kappa^0_-}^{\kappa^0_+}\frac{d\kappa}{\kappa^2\sqrt{I_0}} = \frac{4\kappa'\sqrt{(\kappa'^{\;2} -m^2\lambda^2)}}{4m^2\kappa'^2 +\lambda^2Q^2(4m^2+Q^2)}~, \;\;
\]
\be
 \int_{\kappa^0_-}^{\kappa^0_+}\frac{d\kappa}{\kappa \sqrt{I_0}} ={\cal L}_0~,
 \label{int dkappa 0}
\ee
where
\be
{\cal L}_0 =\frac{1}{\sqrt{Q^2(4m^2+Q^2)}}\ln\left(\frac{\kappa'(2m^2+Q^2) +\sqrt{(\kappa'^2-m^2\lambda^2)Q^2(4m^2+Q^2)}}{\kappa'(2m^2+Q^2)
	-\sqrt{(\kappa'^2-m^2\lambda^2)Q^2(4m^2+Q^2)}} \right)~. \label{L0}
\ee
It gives for $W_i$ (\ref{Wi 1}) in the region $m^2 \gg \kappa'> m\lambda$
\be
W_g = \frac{2m^2-Q^2}{2}\frac{dw}{d\kappa'}~, \;\; W_l = m^2\frac{4m^2+Q^2}{4}\frac{dw}{d\kappa'}~, \;\; \label{dwdkappa'}
\ee
where $m dw/d\kappa'$ is the  spectral probability density for  soft photon emission  with account of photon mass in the rest frame of the final electron,
\be
\frac{dw}{d\kappa'} =\frac{\alpha}{\pi}\frac{1}{\kappa'}
\Bigg[(2m^2+Q^2){\cal L}_0-
-\frac{\sqrt{\kappa'^2-m^2\lambda^2}}{\kappa'}
-\frac{4m^2\kappa'\sqrt{\kappa'^2-m^2\lambda^2}}{4m^2\kappa'^2 +\lambda^2Q^2(4m^2+Q^2)} \Bigg]~.\label{dwdkappaprime}
\ee
Using (\ref{F1}),  (\ref{F2}) and (\ref{dwdkappa'}), one obtains
\be
F_2(x,Q^2) = 2F_1(x,Q^2) = \frac{Q^2}{2}\frac{dw}{d\kappa'}~,
\ee
so that  (\ref{d sigma d q2 dx}) gives for the soft photon emission cross section
\be
\frac{d\sigma^{\gamma}_{soft}}{dQ^2dx} =\frac{d\sigma^{B}}{dQ^2}\frac{Q^2}{2}\frac{dw}{d\kappa'}~.\label{dsoft}
\ee
Integration (\ref{dsoft}) over  the region  $\kappa_0> \kappa' >m\lambda$ ($1-m\lambda/Q^2>x>1-2\kappa_0/Q^2~, \;\; dx=-2d\kappa'/Q^2$) at $\kappa_0\ll m^2, \;\kappa_0\ll Q^2$ provides at $\lambda \rightarrow 0$
\[
\frac{d\sigma_{soft}^{\gamma}}{dQ^2} =\frac{d\sigma_{B}}{dQ^2} \delta^e_{soft}~,
\]
\[
\delta^e_{soft} = \frac{\alpha}{\pi}\Bigg[\frac{1}{\beta} \Bigg(\ln\left(\frac{(1+\beta)}{(1-\beta)}\right)\left(\ln\left(\frac{2\kappa_0}{m\lambda}\right)  +\frac12\right) +\frac12 \text{Li}_2\left(1-\frac{(1+\beta)}{(1-\beta)}\right)
\]
\be
-\frac12 \text{Li}_2\left(1-\frac{(1-\beta)}{(1+\beta)}\right)\Bigg)
-2\ln\left(\frac{2\kappa_0}{m\lambda}\right) +1 \Bigg]~,
\ee
where $\beta$ is given by (\ref{beta}).

At $Q^2\gg m^2$ one has
\be
\delta^e_{soft} = \frac{\alpha}{\pi}\Bigg[2 \ln\left(\frac{2\kappa_0}{m\lambda}\right)  \Big(\ln\left(\frac{Q^2}{m^2}\right)-1 \Big)  -\ln^2\left(\frac{Q^2}{m^2}\right) + \ln\left(\frac{Q^2}{m^2}\right) - \frac{\pi^2}{6}+1\Bigg]~,
\ee
which together with (\ref{delta vertex}) gives
\be
\delta^{e}_{vertex}+ \delta^e_{soft} = \frac{\alpha}{\pi}\Bigg[2 \ln\left(\frac{2\kappa_0}{m^2}\right)  \Big(\ln\left(\frac{Q^2}{m^2}\right)-1 \Big)  -\frac32\ln^2\left(\frac{Q^2}{m^2}\right) + \frac52\ln\left(\frac{Q^2}{m^2}\right) -1\Bigg]~.\label{delta virtsoft}
\ee
As it should be, the dependence on $\lambda$ disappeared in the sum of corrections  from elastic scattering and soft photon emission.

To find the contribution of real photons with $\kappa'>\kappa_0$ for arbitrary $Q^2$ is not so easy. In the following we  restrict ourselves to considering  the case $Q^2\gg m^2$. In this case    it is possible  to introduce the intermediate scale $\kappa_1$, such that $Q^2\gg \kappa_1\gg  m^2$, and calculate the contributions of the regions $\kappa' < \kappa_1$ and $\kappa' > \kappa_1$, simplifying the integrands in them as it is described in Appendix B.   In the sum of these contributions
dependence on the intermediate scale disappears (see (\ref{delta hard one loop})). The remaining dependence on the boundary $\kappa_0$ between soft and hard emission  vanishes  in the  sum of  the correction (\ref{delta hard one loop}) due to  the hard emission   with  $\delta^{e}_{vertex}+ \delta^e_{soft}$ (\ref{delta virtsoft}), so that for the total  correction $\delta^{e}$  one has  in the one-loop approximation  at $Q^2\gg m^2$
\[
\delta^{e}_{one-loop} =  \frac{\alpha}{2\pi}\Bigg[ \big(x_-+\frac{x_-^2}{2}+2\ln(1-x_-)\big) \ln\left(\frac{Q^2}{m^2}\right)
\]
\[
-2\text{Li}_2(x_-)-2\ln x_-\ln(1-x_-) - \ln^2(1-x_-)
-\frac{x_-}{2}(2+x_-)\ln x_- -(2+x_-+\frac{x^2_-}{2})\ln(1-x_-)
\]
\[
  -\frac32 x_- -x_-^2 +\frac{Q^2}{4(pl)}\frac{{Q^2}G_M^2 +4M^2G_E^2}{(4M^2+Q^2)R(1, Q^2)}\Bigg(
(2\ln x_-
\]
\[
-(1-x_-^2))\ln\left(\frac{Q^2}{m^2}\right)-\ln^2 x_--\frac{\pi^2}{3}+2 \text{Li}_2(x_-)+(1-x_-)(3+2x_-)+(1-x_-^2)\ln(x_-(1-x_-))\Bigg)
\]
\[
+\frac{Q^2}{16(pl)^2}\frac{Q^2(Q^2+2M^2)G_M^2 -8M^4G_E^2}{(4M^2+Q^2)R(1, Q^2)} \Bigg( \big((1-x_-^2)\frac{(2+x_-)}{x_-}-2\ln x_- \big)\ln\left(\frac{Q^2}{m^2}\right)+\frac{\pi^2}{3}
\]
\[
-2 \text{Li}_2(x_-) +\ln^2 x_- -(1-x_-^2)\frac{(2+x_-)}{x_-}\ln(1-x_-)-\left(\frac{2}{x_-}+1 -2x_--x_-^2\right)\ln x_-
\]
\be
-(1-x_-)(\frac{1}{x_-}+5+2 x_-)\Bigg)+\frac{Q^2}{4(pl)^2}\frac{Q^2G_M^2 -2M^2G_E^2}{R(1, Q^2)}\ln x_- \Bigg]~.\label{delta one loop}
\ee
The only approximation used here is $Q^2\gg m^2$.

The correction is strongly simplified at small $x_-$ (i.e. at small $Q^2$),  when we have
\[
\delta^{e}_{one-loop} = \frac{\alpha}{2\pi(1+2\rho x_-)}\Bigg[\Big(-2x_- -x_-^2\big(1-\ln x_-+\rho (3-2\ln x_-)\big)\Big)\ln\left(\frac{Q^2}{m^2}\right) + x_-(2\ln x_- -1)
\]
\be
+ x_-^2\Big( -\frac12\ln^2 x_- - \ln x_--\frac{\pi^2}{6} +\frac12 +\rho\big( -\ln^2 x_-+3\ln x_--\frac{\pi^2}{3}\big)\Big)\Bigg]~, \label{delta one-loop at small x}
\ee
where $\rho = (pl)/M^2$.  As we can see, only the terms with power smallness in $x_-$ remain in the full correction.  The terms that do not have such smallness cancel out not only if they are strengthened by powers of $ \ln (Q ^ 2 / m ^ 2) $, but also without such strengthening, as it was noted in \cite{Fadin:2018dwp}.

\section{Parton picture}

In the parton picture the structure functions $F_i(x, Q^2)$  are expressed through parton distributions. In the leading logarithmic approximation (LLA)
\be
F_2(x, Q^2)= 2xF_1(x, Q^2)= x(f^e_e(x, Q^2) + f^{\bar e}_e(x, Q^2))~, \;\;
\ee
where $f^e_e(x, Q^2)$ and  $f^{\bar e}_e(x, Q^2)$ are the electron and positron distributions in the initial electron and the first equality is the Callan-Gross relation \cite{Callan:1969uq}, arising from the fact that the partons have spin 1/2.

In the LLA the parton distributions can be calculated using the  equations \cite{Lipatov:1974qm, Altarelli:1977zs}
\be
\frac{df^a_e(x,Q^2)}{d\ln
	Q^2}=\frac{\alpha(Q^2)}{2\pi}\sum_{b} \int_x^1
\frac{dz}{z}P^a_b(\frac{x}{z})f^b_e(z,Q^2),\label{evolution	equations}
\ee
where $a, b= e, \bar e, \gamma$ , $P^a_b({z})$  are the splitting functions,
\[
P^\gamma_{e}(z)=P^\gamma_{\bar e}(z) =\frac{1+(1-z)^2}{z},\;\;
P^e_\gamma(z)=P^{\bar e}_\gamma(z)=  z^2+(1-z)^2~,\label{splitting functions}
\]
\be
P^e_e(z)=P^{\bar e}_{\bar e}(z)=\frac{1+z^2}{(1-z)_+}+\frac32\delta(1-z)~.
\ee
Here the generalized function $\frac{1}{(1-z)_+}$ is defined by the relation
\be
\int_0^1\frac{f(z)}{(1-z)_+}dz=\int_0^1\frac{f(z)-f(1)}{(1-z)}dz~.
\label{prescription +}
\ee
The evolution equations must be complemented by the initial conditions, which can be taken as
\be
f^a_e(x,m^2) = \delta^a_{e}\delta(1-x)~. \label{initial conditions}
\ee
Presenting parton distributions as the sum of the distributions of valence (v) and sea (s) partons in electron
\be
f^e_e(x,Q^2) = f_e^v(x,Q^2)+  f_e^s(x,Q^2)~,\quad f^{\bar e}_e(x,Q^2) = f^s_e(x,Q^2)~,
\ee
we obtain that these distributions obey  the equations
\be
\frac{df^v_e(x,Q^2)}{d\ln
	Q^2}=\frac{\alpha(Q^2)}{2\pi} \int_x^1
\frac{dz}{z}P^e_e(\frac{x}{z})f^v_e(z,Q^2),\;\; \label{fv equation}
\ee
\be
\frac{df^s_e(x,Q^2)}{d\ln
	Q^2}=\frac{\alpha(Q^2)}{2\pi} \int_x^1
\frac{dz}{z}\bigg(P^e_e(\frac{x}{z})f^s_e(z,Q^2) +  P^e_\gamma(\frac{x}{z})f^\gamma_e(z,Q^2)\bigg),\;\;\label{fs equation}
\ee
with initial
\be
f^v_e(x,m^2) = \delta(1-x)~, \;\;  f^s_e(x,m^2) = 0~ \label{initial conditions 1}
\ee
and charge conservation
\be
\int_0^1f^v_e(x,Q^2) dx =1 \label{int fv=1}
\ee
conditions.
Writing with the two-loop accuracy
\be
f^v_e(x,Q^2)  = \delta(1-x) + \frac{\alpha}{2\pi}L\; V_1(x)  + \left(\frac{\alpha}{2\pi}\right)^2 \frac{L^2}{2}\; V_2(x)~, \;\;\label{fv}
\ee
\be
f^s_e(x,Q^2)  =  \left(\frac{\alpha}{2\pi}\right)^2 \frac{L^2}{2}S_2(x)~, \label{fs}
\ee
where $L=\ln(Q^2/m^2)$~,
and taking into account  in $\alpha(Q^2)$ only  the one-loop correction  coming from vacuum polarisation by electrons
\be
\alpha(Q^2) = \alpha + \frac{\alpha^2}{3\pi}L~, \;\;
\ee
we have
\be
V_1(x)= P^e_e(x)=\frac{1+x^2}{(1-x)_+}+\frac32\delta(1-x)~, \label{V1}\;\;
\ee
\[
V_2(x)= \frac{2}{3}\, P^e_e(x)
+ \int_{x}^{1} \frac{d z}{z} P_{e}^{e}\left(\frac{x}{z}\right) P_{e}^{e}\left(z\right) =\frac{2}{3}P^e_e(x)  +8\left(\frac{\ln(1-x)}{(1-x)}\right)_+
\]
\be
+ \frac{1+4x+x^2}{(1-x)}_+ -\frac{1+3x^2}{(1-x)} \ln x - 4(1+x)\ln(1-x) -\bigg(\frac{2\pi^2}{3}-\frac{9}{4}\bigg)\delta(1-x)~, \label{V2}
\ee
\be
S_2(x) = \int_{x}^{1} \frac{d z}{z} P_{e}^{\gamma}\left(\frac{x}{z}\right)
P_{\gamma}^{e}\left(z\right) = 2(1+x)\ln x +1-x +\frac{4(1-x^3)}{3x}~, \label{S2}
\ee
where   the generalized function $\left(\frac{\ln(1-z)}{(1-z)}\right)_+$ are defined by the relation  analogous to (\ref{prescription +})
\be
\int_0^1\left(\frac{\ln(1-z)}{(1-z)}\right)_+ f(z)dz=\int_0^1\left(\frac{\ln(1-z)}{(1-z)}\right)(f(z)-f(1))dz~.
\label{prescription ++}
\ee
The coefficients of the delta-function terms in (\ref{V1}), (\ref{V2})  are  determined by the
requirements
\be
\int_0^1 dx V_i(x) =0~ \label{int vi=0}
\ee
following from the charge conservation condition (\ref{int fv=1}).

Writing the  cross section (\ref{d sigma d q2 dx}) at $Q^2\gg m^2, \;\; F_2(x, Q^2) = 2x  F_1(x, Q^2)  $ as
\be
\frac{d\sigma}{dQ^2 dx} = \frac{4\pi\left(\alpha(Q^2)\right)^2}{Q^4}\frac{F_2(x, Q^2)}{x} R(x, Q^2)~,
\ee
where  $R(x, Q^2)$ is given by (\ref{Rx}),
we have for the radiative correction $\delta^e$ in the  leading logarithmic approximation
\be
\delta^e_{LLA}  = \int_{x_-}^1 dx\frac{R(x, Q^2)}{R(1, Q^2)}(f^v_e(x,Q^2) + 2f^s_e(x,Q^2)) -1~,
\ee
This representation permits to find the radiative correction $\delta^e$ in any order of perturbation theory.

With  the two-loop accuracy   $f^v_e(x,Q^2)$ and  $f^s_e(x,Q^2)$ are  given  by Eqs. (\ref{fv}) and  (\ref{fs})
respectively.  Using (\ref{Rx}) and (\ref{int fv=1}), we have
\be
\delta^e_{LLA}   =  - \int_{0}^{x_-} dx f^v_e(x,Q^2) +  \int_{x_-}^1dx \bigg[\bigg(\frac{R(x, Q^2)}{R(1, Q^2)}-1\bigg)f^v_e(x,Q^2)+ 2\frac{R(x, Q^2)}{R(1, Q^2)}f^s_e(x,Q^2)\bigg]~.  \label{delta e l}
\ee

In the one-loop approximation only $f^v_e(x,Q^2)$ does contribute. Simple integration gives % checked
\[
\delta^e_{one-loop, LLA} = \frac{\alpha}{2\pi} L \Bigg(2\ln(1-x_-) +x_- + \frac{x_-^2}{2}   + \frac{Q^2}{2(pl)}\frac{{Q^2}G_M^2 +4M^2G_E^2}{(4M^2+Q^2)R(1, Q^2)}\left(\ln x_--\frac{1-x_-^2}{2}\right)
\]
\be
+ \frac{Q^2}{8(pl)^2}\frac{Q^2(2M^2 + Q^2)G_M^2 -8M^4G_E^2}{(4M^2+Q^2)R(1, Q^2)}\left(-\ln x_- +(1-x_-)\left(\frac{1}{x_-}+\frac32 +\frac{x_-}{2}\right)\right)\Bigg) ~,
\ee
in accordance with (\ref{delta one loop}).

The two-loop correction contains contributions of both $f_e^v$ and $f_e^s$. Using (\ref{fv})-(\ref{S2}) and (\ref{Rx}), one obtains from (\ref{delta e l}) the two-loop contribution in the form
\[
\delta^e_{two-loop, LLA} = \left(\frac{\alpha}{2\pi}\right)^2 \frac{L^2}{2}\Bigg[-\int_0^{x_-} dx V_2(x)+ \frac{{Q^2}G_M^2 +4M^2G_E^2}{(4M^2+Q^2)R(1, Q^2)}\int^1_{x_-} dx \bigg(2S_2(x)-\frac{Q^2}{2(pl)}\bigg(\frac{1-x}{x}
\]
\be
\times V_2(x)+\frac{2}{x} S_2(x)\bigg)\bigg)+ \frac{Q^2}{8(pl)^2}\frac{Q^2(2M^2 + Q^2)G_M^2 -8M^4G_E^2}{(4M^2+Q^2)R(1, Q^2)}\int^1_{x_-} dx \big(\frac{1-x^2}{x^2}V_2(x)+\frac{2}{x^2}S_2(x)\big)\bigg)
\Bigg] ~.   \label{delta 2 general}
\ee
Elementary integration gives
\[
\int^{x_-}_0 dx V_2(x) = 4\text{Li}_2(x_-)-4\ln(1-x_-)\ln\frac{(1-x_-)}{x_-} -(\frac{4}{3}+4x_- +2x^2_-)\ln(1-x_-)
\]
\be
+3x_-(1+\frac{x_-}{2})\ln x_- -\frac83 x_- - \frac{7}{12} x^2_-~,  \label{v2-1}
\ee
\[
\int_{x_-}^1 dx\left(\frac{1}{x}-1\right) V_2(x) = 4\text{Li}_2(x_-)-\frac23 \pi^2 +\frac12\ln^2x_-+2(1-x_-^2)\ln(1-x_-)
\]
\be
-(\frac53-\frac32 x_-^2)\ln x_- +\frac{1}{12}(1-x_-)(31+7x_-) ~, \label{v2-2}
\ee
\[
\int_{x_-}^1 dx\left(\frac{1}{x^2}-1\right) V_2(x) = 4\text{Li}_2(x_-)-\frac23 \pi^2 +\frac12\ln^2x_-   +2(1-x_-^2)(\frac{2}{x_-} +1)\ln(1-x_-)
\]
\be
- (\frac{1}{x_-}+\frac53 - 3 x_- -\frac32 x_-^2)\ln x_- + (1-x_-)(\frac{2}{3x_-}+\frac{13}{4}+\frac{7}{12}x_-)~,  \label{v2-3}
\ee
\be
\int_{x_-}^1 dx S_2(x) = -(\frac43 +2x_- +x^2_-)\ln x_- - \frac19(1-x_-)(22+13 x_- +4x_-^2)
~,  \label{s2-1}
\ee
\be
\int_{x_-}^1 dx \frac{S_2(x)}{x} = - \ln^2x_- - ({2}{x_-} +1)\ln x_-  + \frac{(1-x_-)}{3}(\frac{4}{x_-}-11 -2x_-)~,
\label{s2-2}
\ee
\be
\int_{x_-}^1 dx \frac{S_2(x)}{x^2} = - \ln^2x_- + (\frac{2}{x_-} +1)\ln x_-  + \frac{(1-x_-)}{3}(\frac{2}{x_-^2}+\frac{11}{x_-}-4)~,  \label{s2-3}
\ee
so that % checked
\[
\delta^e_{two-loop, LLA} = \left(\frac{\alpha}{2\pi}\right)^2 \frac{L^2}{2}\Bigg[-4\text{Li}_2(x_-)+4\ln(1-x_-)\ln\frac{(1-x_-)}{x_-} +(\frac{4}{3}+4x_- +2x^2_-)\ln(1-x_-)
\]
\[
-3x_-(1+\frac{x_-}{2})\ln x_- +\frac83 x_- + \frac{7}{12} x^2_-)
\]
\[
+ \frac{{Q^2}G_M^2 +4M^2G_E^2}{(4M^2+Q^2)R(1, Q^2)}\Bigg(-2(\frac43 +2x_- +x^2_-)\ln x_- - \frac29(1-x_-)(22+13 x_- +4x_-^2)
\]
\[
-\frac{Q^2}{2(pl)}\bigg(4\text{Li}_2(x_-)-\frac23 \pi^2 -\frac32\ln^2x_-   +2(1-x_-^2)\ln(1-x_-)
\]
\[
 -(\frac{11}{3}+4x_--\frac32x^2_-)\ln x_- +(1-x_-)(\frac{8}{3x_-}-\frac{19}{4}-\frac34 x_-)\bigg)\Bigg)
\]
\[
+ \frac{Q^2}{8(pl)^2}\frac{Q^2(2M^2 + Q^2)G_M^2 -8M^4G_E^2}{(4M^2+Q^2)R(1, Q^2)}\bigg(4\text{Li}_2(x_-)-\frac23 \pi^2 -\frac32\ln^2x_-   +2(1-x_-^2)(\frac{2}{x_-} +1)\ln(1-x_-)
\]
\be
+ (\frac{3}{x_-}+\frac13 +3 x_- +\frac32 x_-^2)\ln x_- + (1-x_-)(\frac{4}{3x^2_-}+\frac{8}{x_-}+\frac{7}{12}(1+x_-))
\bigg)\Bigg] ~.   \label{delta 2 full}
\ee
Note that at small momentum transfer, i.e. at small $x_-$,  the valence quark contribution is suppressed as well as in the one loop due to the charge conservation requirement  (\ref{int vi=0}). It is not so for the sea quark contribution \cite{Fadin:2018dwp}.  The sea quark distribution is singular at $x=0$  and  the lower limit $x_0$ of the  integration in (\ref{s2-1}) can not be taken equal to 0.  Therefore the two-loop correction is not suppressed at small momentum transfer for  experimental conditions at which production of electron-positron pairs is not forbidden. For such conditions we have at $x_- \ll 1$
\be
\delta^e_{two-loop, LLA} = \left(\frac{\alpha}{2\pi}\right)^2 \frac{L^2}{2}\bigg[ -\frac83 \ln x_-
-\frac{44}{9}- \frac{4}{3(1+2\rho x_-)}\bigg] ~.
\label{delta 2 at small Q}
\ee
At first glance it seems that more preferable are the conditions at which production  of electron-positron pairs is forbidden. In this case   $f_e^s$ must be omitted   in Eq.~(\ref{delta e l}), and   the term with  $ P_e^e(x)$ must be omitted in its expression for $V_2(x)$ in (\ref{V2}).  However, this is not the whole truth. The term with  $ P_e^e(x)$  in $V_2(x)$  meets  contributions from  not only real, but also virtual pairs, which can not be suppressed, and therefore their contribution must be restored. It means that  the term
\be
\left(\frac{\alpha}{\pi}\right)^2 \bigg[-\frac{1}{36}L^3 +\frac{19}{72}L^2\bigg]\delta(1-x)
\ee
must be added to  $f_e^v$ (see for details \cite{Kuraev:1985wb}). Therefore, in this case
\[
\delta^e_{two-loop, LLA} = \left(\frac{\alpha}{2\pi}\right)^2 \frac{L^2}{2}\Bigg[-\frac29 L +\frac{19}{9} -4\text{Li}_2(x_-)+4\ln(1-x_-)\ln\frac{(1-x_-)}{x_-}
\]
\[
+2x_-(2 + x_-)\ln(1-x_-) -3x_-(1+\frac{x_-}{2})\ln x_- +2 x_- + \frac{1}{4} x^2_-
\]
\[
-\frac{Q^2}{2(pl)} \frac{{Q^2}G_M^2 +4M^2G_E^2}{(4M^2+Q^2)R(1, Q^2)}
\bigg(4\text{Li}_2(x_-)-\frac23 \pi^2 +\frac12\ln^2x_-   +2(1-x_-^2)\ln(1-x_-)
\]
\[
-(1-\frac32x^2_-)\ln x_- +\frac14(1-x_-)(9+x_-)\bigg)
\]
\[
+ \frac{Q^2}{8(pl)^2}\frac{Q^2(2M^2 + Q^2)G_M^2 -8M^4G_E^2}{(4M^2+Q^2)R(1, Q^2)}\bigg(4\text{Li}_2(x_-)-\frac23 \pi^2 +\frac12\ln^2x_-   +2(1-x_-^2)(\frac{2}{x_-} +1)\ln(1-x_-)
\]
\be
- (\frac{1}{x_-}+1 -3 x_- +\frac32 x_-^2)\ln x_- + \frac14(1-x_-)(9+x_-)
\bigg)\Bigg] ~.   \label{delta 2 suppressed}
\ee
The relative magnitude of the corrections (\ref{delta 2 general}) and (\ref{delta 2 suppressed}) depends on energy and momentum transfer.

\section{Conclusion}
As it was shown in \cite{Fadin:2018dwp},  the setting of the experiment with recoil proton detection  suggested by A.A. Vorobev  \cite{Vorobev} for measurement of proton radius,  has an interesting feature -- cancellation of main radiative corrections.  Here we calculated radiative corrections to the cross section of electron-proton scattering for experiments of this kind in a wide range of kinematic parameters using the method of   structure functions and parton distributions. We calculated the  one-loop corrections  due to electron interaction for momentum transfer $Q$ limited only by the requirement $Q\gg m$, $m$ being the electron mass, and proved that when  at small $Q$   the cancellation  of the virtual and real radiative corrections has a power accuracy.

In the two-loop approximation we calculated these corrections with logarithmic accuracy, again for momentum transfer $Q$ limited only by the requirement $Q\gg m$, using the parton distribution  method \cite{Gribov:1972ri}-\cite{Altarelli:1977zs} developed for the theoretical description of  the deep inelastic electron-proton scattering and adopted  in \cite{Kuraev:1985wb} for calculation of radiative corrections in QED.  We calculated the radiation corrections both for such an experiment setup when the production of additional electron-positron pairs is allowed, and for such when it is forbidden.

\section*{ Appendix A} \label{section: appendix}

\def\theequation{A.\arabic{equation}}

The vacuum polarisation  ${\cal P}(q^2)$ contains lepton (electron, muon, $\tau$-lepton ) and hadron contributions:
\be
{\cal P}(q^2) =  {\cal P}_e(q^2) + {\cal P}_\mu(q^2) +{\cal P}_\tau(q^2) + {\cal P}_h(q^2).
\ee
One-loop lepton contribution $\mathcal{P}^{(1)}_{l}(q^2), \;\;( l=e, \mu, \tau)$ is well known (see, for example, \cite{BLP}):
\be
{\cal P}^{(1)}_{l}(q^2) = \frac{\alpha}{\pi}\left(\frac{1}{3}\,\sqrt{1-\frac{4\,m_{l}^2}{q^2}}\,\left(1+ \frac{2\,m_{l}^2}{q^2}\right)\,\ln\left(\frac{\sqrt{1-\frac{4m_{l}^2}{q^2}}+1}{\sqrt{1-\frac{4m_{l}^2}{q^2}}-1}\right)-\frac{4\,m_{l}^2}{3\,q^2} - \frac{5}{9}\right).
\ee
At $Q^2 =-q^2 \gg 4m_l^2$
\be
{\cal P}^{(1)}_{l}(q^2) =\frac{\alpha}{3\pi}\left(\ln\left(\frac{Q^2}{m_l^2}\right) - \frac53\right)~,
\ee
and at $Q^2 =-q^2 \ll 4m_l^2$
\be
{\cal P}_{l}(q^2) = \frac{\alpha}{15\pi}\frac{Q^2}{m_l^2}~.
\ee
The lepton contributions are known also  in higher orders of perturbation theory  (see, for example, \cite{DeRafael:1974iv, Aoyama:2020ynm}). For us  it is enough  to know that the two-loop contribution contains only the first degree of $\ln\left(\frac{Q^2}{m_l^2}\right)$.

The hadron contribution ${\cal P}_h(q^2)$ is expressed   in terms of the total cross section of  one-photon electron-positron pair annihilation  into hadrons
\be
{\cal P}_h(q^2) = \frac{q^2}{4\pi^2\alpha}\int_{4m^2_{\pi}}^\infty ds\frac{\sigma_{e^+e^ \rightarrow hadrons}(s)}{s-q^2}~.
\ee
This contribution is small compared with $\alpha/\pi$ at $Q^2 < 4m^2_{\pi}$,  becomes of order of $\alpha/\pi$ only at  $Q^2\sim 4m^2_{\pi}$  and then grows logarithmically with $Q^2$.  Recent review is given in  \cite{Aoyama:2020ynm}.

\section*{ Appendix B} \label{section: appendix B}

\def\theequation{B.\arabic{equation}}

To find the contribution $\delta^e_{hard}$  of the one-photon emission  with $\kappa'>\kappa_0$ to the radiative correction $\delta^e$  at $Q^2\gg m^2$  it is convenient to introduce the intermediate scale $\kappa_1$ such that  $Q^2\gg \kappa_1\gg  m^2$.
In the region $\kappa_1>\kappa'>\kappa_0$ one can put
\be
W_1=-\frac{\alpha}{2\pi}\Bigg[ \frac{Q^2}{\kappa'} \Big(\ln\left(\frac{Q^4}{m^2(m^2+2\kappa')}\right)-2 \Big)  +\frac{Q^2\kappa'}{(m^2+2\kappa')^2}\Bigg]~,
\ee
\be
W_2=0~,\;\;   F_1= -\frac12 W_1~,\;\;F_2=2F_1~.
\ee
Therefore in this region  we have from Eqs. (\ref{d sigma dQ^2 dx})--(\ref{d sigma B d q 2})
\be
\frac{d\sigma^{\gamma}}{dQ^2 dx } =  - W_1\frac{d\sigma_{B}}{dQ^2}~.
\ee
Using that in this region it is possible to put $\kappa' = Q^2(1-x)/2$, it is easy to obtain the part $\delta^{(1)}_{hard}$   of the correction  $\delta^e_{hard}$, defined by Eqs. (\ref{delta el}) and  (\ref{delta e}),    from this region:
\[
\delta^{(1)}_{hard} = \frac{\alpha}{\pi}\int_{\kappa_0}^{\kappa_1}\frac{d\kappa'}{\kappa'}\Bigg[2 \Big(\ln\left(\frac{Q^2}{m^2}\right)-1 \Big) - \ln\left(\frac{m^2+2\kappa'}{m^2}\right) + \frac{\kappa'^2}{(m^2+2\kappa')^2} \Bigg]
\]
\be
= \frac{\alpha}{\pi}\Bigg[2 \ln\left(\frac{\kappa_1}{\kappa_0}\right)  \Big(\ln\left(\frac{Q^2}{m^2}\right)-1 \Big)  -\frac12\ln^2\left(\frac{2\kappa_1}{m^2}\right) + \frac14\Big(\ln\left(\frac{2\kappa_1}{m^2}\right)-1\Big) -\frac{\pi^2}{6}\Bigg]~,
\ee
Using (\ref{delta virtsoft}), we obtain
\[
\delta^{e}_{vert}+ \delta_{soft}+\delta^{(1)}_{hard} =\frac{\alpha}{\pi}\Bigg[2 \ln\left(\frac{2\kappa_1}{m^2}\right)  \Big(\ln\left(\frac{Q^2}{m^2}\right)-1 \Big)  -\frac32\ln^2\left(\frac{Q^2}{m^2}\right)
\]
\be
+ \frac52\ln\left(\frac{Q^2}{m^2}\right)  -\frac12\ln^2\left(\frac{2\kappa_1}{m^2}\right) + \frac14 \ln\left(\frac{2\kappa_1}{m^2}\right) -\frac{\pi^2}{6} -\frac54\Bigg]~.
\ee
In the region $\kappa_{max}>\kappa'>\kappa_1$,  i.e.  $1-\frac{2\kappa_1}{Q^2}>x>x_{-}$ at $Q^2\gg m^2$ one can put
\be
W_1 = -\frac{\alpha}{2\pi}\Big( \frac{1+x^2}{1-x}\ln\left(\frac{Q^2}{m^2x(1-x)}\right) +\frac{1-8x}{2(1-x)} \Big)~,
\ee
\be
W_2=\frac{\alpha}{2\pi}\frac{Q^2}{4x}~,
\ee
and
\be
F_1 = \frac12\left(\frac{4x^2\;W_2}{Q^2}-W_1\right)~,
\ee
\be
F_2 =x\left(\frac{12x^2}{Q^2}\;W_2-W_1\right)~ =2x F_1 +\frac{\alpha}{\pi}x^2~.
\ee
As it is seen, the Callan-Gross relation \cite{Callan:1969uq} is violated in this region.  It could be expected, since this relation is valid only in the collinear approximation.

Using Eqs. (\ref{d sigma dQ^2 dx})-(\ref{d sigma B d q 2}), we have for the part $\delta^{(2)}_{hard}$   of the  correction $\delta^e_{hard}$  from this region:
\[
\delta^{(2)}_{hard} = \int_{x_-}^{1-2\frac{\kappa_1}{Q^2}}dx \bigg[\frac{F_2(x, Q^2)}{x}  -\frac{Q^2}{2(pl)}\frac{Q^2G_M^2 +4M^2G_E^2}{(4M^2+Q^2)R(1, Q^2)}\frac{F_2(x, Q^2)(1-x)}{x^2}
\]
\[
 +\frac{Q^2}{8(pl)^2}\Bigg(\frac{Q^2(Q^2+2M^2)G_M^2 -8M^4G_E^2}{(4M^2+Q^2)R(1, Q^2)} \Bigg)\frac{F_2(x, Q^2)(1-x^2)}{x^3}
\]
\be
-\frac{Q^2}{8(pl)^2}\frac{(Q^2G_M^2 -2M^2G_E^2)}{R(1, Q^2)}\frac{(F_2(x, Q^2)-2xF_1(x, Q^2))}{x^3}\bigg] \label{integral for delta 12 hard}~.
\ee
Note that in the integral (\ref{integral for delta 12 hard}) the upper limit  can be set equal to 1 in all terms except the first one. It gives
\[
\delta^{(2)}_{hard} = \frac{\alpha}{2\pi}\Bigg[ \Bigg(3\ln\left(\frac{Q^2}{m^2}\right) -4\ln\left(\frac{2\kappa_1}{m^2}\right)-5+x_-+\frac{x_-^2}{2}+2\ln(1-x_-)\Bigg) \ln\left(\frac{Q^2}{m^2}\right)
\]
\[
+\ln^2\left(\frac{2\kappa_1}{m^2}\right) +\frac72\ln\left(\frac{2\kappa_1}{m^2(1-x_-)}\right)-\ln^2(1-x_-)
+2\text{Li}_2(1-x_-)+\frac52
\]
\[
-\frac{x_-}{2}(2+x_-)\ln x_- +\frac{(1-x_-)}{2}(3+x_-)\ln(1-x_-) -\frac{x_-}{2}(3+2x_-)\Bigg]
\]
\[
+\frac{Q^2}{4(pl)}\frac{Q^2G_M^2 +4M^2G_E^2}{4M^2+Q^2}\Bigg(
(2\ln x_--(1-x_-^2))\ln\left(\frac{Q^2}{m^2}\right)-\ln^2 x_--\frac{\pi^2}{3}+2 \text{Li}_2(x_-)
\]
\[
+(1-x_-)(3+2x_-)+(1-x_-^2)\ln(x_-(1-x_-))\Bigg) +\frac{Q^2}{16(pl)^2}\Bigg(\frac{Q^2(Q^2+2M^2)G_M^2 -8M^4G_E^2}{4M^2+Q^2} \Bigg)
\]
\[
\times\Bigg( \big((1-x_-^2)\frac{(2+x_-)}{x_-}-2\ln x_- \big)\ln\left(\frac{Q^2}{m^2}\right)+\frac{\pi^2}{3}-2 \text{Li}_2(x_-) +\ln^2 x_- -(1-x_-^2)\frac{(2+x_-)}{x_-}\ln(1-x_-)
\]
\be
-\left(\frac{2}{x_-}+1 -2x_--x_-^2\right)\ln x_-  -(1-x_-)(\frac{1}{x_-}+5+2 x_-)\Bigg)
+\frac{Q^2}{4(pl)^2}(Q^2G_M^2 -2M^2G_E^2)\ln x_-  \Bigg]~.
\ee
The intermediate parameter $\kappa_1$ disappears in the sum $\delta_{hard}^e =  \delta^{(2)}_{hard}+ \delta^{(1)}_{hard}$:
\[
\delta^{e}_{hard} =
  \frac{\alpha}{2\pi}\Bigg[ \Big(3\ln\left(\frac{Q^2}{m^2}\right) -4\ln\left(\frac{2\kappa_0}{m^2}\right)-5+x_-+\frac{x_-^2}{2}+2\ln(1-x_-)\Bigg) \ln\left(\frac{Q^2}{m^2}\right)
 \]
 \[
  +4\ln\left(\frac{2\kappa_0}{m^2}\right)-\ln^2(1-x_-)
 +2\text{Li}_2(1-x_-)-\frac{x_-}{2}(2+x_-)\ln x_- -(2+x_-+\frac{x^2_-}{2})\ln(1-x_-)
 \]
 \[
 -\frac{\pi^2}{3}+2  -\frac32 x_- -x_-^2 +\frac{Q^2}{4(pl)}\frac{{Q^2}G_M^2 +4M^2G_E^2}{(4M^2+Q^2)R(1, Q^2)}\Bigg(
(2\ln x_-
\]
\[
-(1-x_-^2))\ln\left(\frac{Q^2}{m^2}\right)-\ln^2 x_--\frac{\pi^2}{3}+2 \text{Li}_2(x_-)+(1-x_-)(3+2x_-)+(1-x_-^2)\ln(x_-(1-x_-))\Bigg)
\]
\[
+\frac{Q^2}{16(pl)^2}\frac{Q^2(Q^2+2M^2)G_M^2 -8M^4G_E^2}{(4M^2+Q^2)R(1, Q^2)} \Bigg( \big((1-x_-^2)\frac{(2+x_-)}{x_-}-2\ln x_- \big)\ln\left(\frac{Q^2}{m^2}\right)+\frac{\pi^2}{3}
\]
\[
-2 \text{Li}_2(x_-) +\ln^2 x_- -(1-x_-^2)\frac{(2+x_-)}{x_-}\ln(1-x_-)-\left(\frac{2}{x_-}+1 -2x_--x_-^2\right)\ln x_-
\]
\be
-(1-x_-)(\frac{1}{x_-}+5+2 x_-)\Bigg)+\frac{Q^2}{4(pl)^2}\frac{Q^2G_M^2 -2M^2G_E^2}{R(1, Q^2)}\ln x_- \Bigg]~. \label{delta hard one loop}
\ee

\end{document}